\documentclass[aps,floatfix,twocolumn,byrevtex,superscriptaddress]{revtex4-1}
\usepackage{amsmath}
\usepackage{amssymb}
\usepackage{amstext}
\usepackage{amsopn}
\usepackage{amsfonts}
\usepackage{amsxtra}
\usepackage[english]{babel}
\usepackage{graphicx}
\usepackage{float}
\usepackage{bm}
\usepackage{multirow}
\usepackage{dcolumn}
\usepackage{color}
\usepackage{hyperref}

\newcommand{\icm}{\ensuremath{~\textrm{cm}^{-1}}}

\begin{document}

\title{Superconductivity-induced transverse plasma mode and phonon anomaly in the $c$-axis response of the bilayer compound RbCa$_2$Fe$_4$As$_4$F$_2$}

\author{B. Xu}
\affiliation{University of Fribourg, Department of Physics and Fribourg Center for Nanomaterials, Chemin du Mus\'{e}e 3, CH-1700 Fribourg, Switzerland}

\author{D. Munzar}
\author{A. Dubroka}
\affiliation{Department of Condensed Matter Physics, Faculty of Science, and Central European Institute of Technology, Masaryk University, Kotl\'{a}\v{r}sk\'{a} 2, 61137 Brno, Czech Republic}

\author{E. Sheveleva}
\author{F. Lyzwa}
\author{P. Marsik}
\affiliation{University of Fribourg, Department of Physics and Fribourg Center for Nanomaterials, Chemin du Mus\'{e}e 3, CH-1700 Fribourg, Switzerland}

\author{C. N. Wang}
\affiliation{Laboratory for Muon Spin Spectroscopy, Paul Scherrer Institute, CH-5232 Villigen PSI, Switzerland}

\author{Z. C. Wang}
\author{G. H. Cao}
\affiliation{Department of Physics and State Key Lab of Silicon Materials, Zhejiang University, Hangzhou 310027, China}

\author{C. Bernhard}
\email[]{christian.bernhard@unifr.ch}
\affiliation{University of Fribourg, Department of Physics and Fribourg Center for Nanomaterials, Chemin du Mus\'{e}e 3, CH-1700 Fribourg, Switzerland}

\date{\today}
%
%

\begin{abstract}
With infrared spectroscopy we studied the out-of-plane ($c$-axis) response of the iron arsenide superconductor ACa$_2$Fe$_4$As$_4$F$_2$ (A = Rb,Cs) which has a bilayer structure similar to the high $T_c$ cuprates YBa$_2$Cu$_3$O$_7$ (YBCO) and Bi$_2$Sr$_2$CaCu$_2$O$_8$ (Bi2212). In analogy to the cuprates, we observe a superconductivity-induced transverse plasma mode (tPM) and a phonon anomaly that are both signatures of local electric field effects that arise from a large difference between the local conductivities in the intra- and inter-bilayer regions. Using a multilayer model developed for the cuprates, we obtain a good description of the $c$-axis response and derive the local conductivities at $T \simeq T_c$ of $\sigma_1^{\mathrm{bl}}(\omega \rightarrow 0) \simeq$ 1\,000 $\Omega^{-1}\mathrm{cm}^{-1}$ and $\sigma_1^{\mathrm{int}}(\omega \rightarrow 0) \simeq$ 15 $\Omega^{-1}\mathrm{cm}^{-1}$, respectively, that are similar to the ones previously found in underdoped YBCO. Different from the cuprates, we find no evidence of a normal state pseudogap in terms of a partial suppression of the low-energy electronic states that sets in already well above $T_c$. There is also no clear sign of an onset of precursor superconducting pairing correlations well above $T_c \simeq$ 30~K. This highlights that the pseudogap and the precursor superconducting pairing well above $T_c$ are unique features of the cuprates with their strong electronic correlations and, for example, not just the result of a strongly anisotropic electronic response due to the layered crystal structure.
\end{abstract}



\maketitle

%
%
\section{Introduction}
\subsection{$c$-axis response of cuprates}
The study of the infrared $c$-axis response of the cuprate high $T_c$ superconductors, like La$_{2-x}$Sr$_x$CuO$_4$ (LSCO) or YBa$_2$Cu$_3$O$_{7-\delta}$ (YBCO), has provided valuable information about their unusual electronic properties in the normal and superconducting states. The incoherent, insulator-like response in the normal state and the emergence of a rather sharp plasma edge in the low-frequency $c$-axis reflectivity below $T_c$ are meanwhile understood in terms of a weak, Josephson-type coupling between the quasi-two-dimensional CuO$_2$ planes. Both the incoherence of the $c$-axis response in the normal state and the very small value of the longitudinal superconducting (SC) plasma frequency below $T_c$ are due to the so-called pseudogap phenomenon, which gives rise to a partial depletion of the low-energy electronic states that sets in far above $T_c$ and persist in the SC state~\cite{Basov2005RMP,Basov1994PRB,Tajima1997PRB,Homes1993PRL,Timusk1999,Puchkov1996,Bernhard2000PRB,Boris2002PRL,Dubroka2010,Dubroka2011PRL}.

The origin of this pseudogap, which governs the unusual electronic properties in the so-called underdoped part of the phase diagram, has been investigated for decades and remains debated. Meanwhile, at least some phenomenological trends are commonly accepted. The pseudogap affects only parts of the Fermi-surface of the cuprates, i.e. it is largest in the so-called anti-nodal region close to (0,$\pi$) or ($\pi$,0) and absent in the nodal region close to the Brillouin-zone diagonal (in the notation of the $d$-wave superconducting order parameter)~\cite{Damascelli2003RMP}. The pseudogap has therefore a strong influence on the $c$-axis conductivity, especially in YBCO for which the matrix-element for the hopping across the spacing layer including the BaO and CuO chain layers displays a similar $k$-space anisotropy, i.e. it is large (small) in the anti-nodal (nodal) regions of the Fermi-surface. Notably, as the pseudogap vanishes on the overdoped side of the phase diagram, the normal state $c$-axis conductivity becomes metal-like, albeit with a much smaller (longitudinal) SC plasma frequency than for the in-plane response.

Several additional, striking features have been observed in the $c$-axis response of the so-called bilayer compounds, like YBCO and Bi-2212~\cite{Bernhard2000PRB,Boris2002PRL,Dubroka2010,Dubroka2011PRL,Uykur2014,Zelezny2001}. They contain pairs of closely spaced CuO$_2$ layers (so-called bilayers) that are separated by an Y or Ca monolayer, respectively. The electronic coupling within the CuO$_2$ bilayer unit is thus considerably stronger than the one across the thicker stacks of BaO$_2$-CuO-BaO$_2$ layers in YBCO and SrO$_2$-BiO$_2$-SrO$_2$ layers in Bi-2212, that separate the CuO$_2$ bilayer units and are in the following denoted as inter-bilayer regions. This leads to large differences between the local conductivities in the intra-bilayer and inter-bilayer regions, $\sigma_1^{\mathrm{bl}}$ and $\sigma_1^{\mathrm{int}}$, and the related plasma frequencies, $\omega_p^{\mathrm{bl}}$ and $\omega_p^{\mathrm{int}}$, respectively. Accordingly, there exists a transverse plasma mode (tPM) that shows up as a peak in the optical conductivity at a frequency $\omega_p^{\mathrm{tr}}$ with $\omega_p^{\mathrm{bl}} > \omega_p^{\mathrm{tr}} > \omega_p^{\mathrm{int}}$ at which the currents in the intra- and inter-bilayer regions oscillate out of phase (the inter-bilayer response may also be insulating with $\omega_p^{\mathrm{int}}$ = 0) ~\cite{VanderMarel1996}. In the incoherent normal state, the tPM is overdamped and barely visible, whereas below $T_c$ it tends to become a pronounced feature as the coherency of the electronic response increases in the superconducting state. The formation of the tPM is typically accompanied by pronounced anomalies of some of the infrared phonon modes. The latter originate from changes of the local electric fields within the intra- and inter-bilayer regions that are caused by dynamical charging of the CuO$_2$ planes ~\cite{Munzar1999}. Accordingly, the anomaly of a phonon mode strongly depends on its frequency and the displacement pattern (i.e. on  which ions participate in the vibration).

In underdoped YBCO the most pronounced SC-induced anomaly occurs for the bending mode at 320\icm\ that involves mainly vibrations along the $c$-axis of the O ions within the CuO$_2$ planes~\cite{Bernhard2000PRB}. The anomalous behaviour of this phonon has been quantitatively reproduced with a relatively simple multilayer model that accounts for the changes of the dynamic local electric fields due to the formation of the tPM~\cite{Munzar1999,Dubroka2010}. A corresponding anomaly of a mode at 190\icm (denoted in the following as Y-mode), which exhibits a sudden narrowing and increases in oscillator strength below $T_c$, can also be qualitatively explained with this multilayer model. A quantitative modelling, however, is difficult, since it requires a precise knowledge of the displacement pattern. Note that the mode involves vibrations of both intra-bilayer ions (Y) and inter-bilayer ones (chain Cu and O)~\cite{Humlicek1993}.

Notably, a detailed study of the temperature and doping dependence of the tPM and the related anomaly of the 320\icm\ phonon mode, has provided evidence for precursor SC pairing correlations that set in at temperatures well above the bulk superconducting critical temperature $T_c$. The bilayer unit acts here as a high frequency resonator with an eigenfrequency of about 300 to 500\icm\ (or 10 to 15~THz) that is sensitive even to short ranged and fast fluctuating SC correlations. Especially the phonon anomalies, due to the local-electric-field effect caused by the tPM, are very sensitive to coherency of the intra-bilayer response and thus the local SC pairing. Moreover, it was shown that large magnetic fields suppress this onset of coherency which excludes alternative explanations in terms of charge-density-wave or antiferromagnetic correlations that would be rather enhanced or hardly affected~\cite{Dubroka2011PRL}.

\subsection{$c$-axis response of pnictides}
The iron arsenide high $T_c$ superconductors, that were discovered in 2008, have a layered structure similar to the one of the cuprates~\cite{Paglione2010,Basov2011}. Their FeAs layers are separated either by monolayers of atoms of alkali, alkaline earth, or rare earth metals or by thicker, insulating layers containing oxygen and/or fluorine. This raises the question of whether similar spectroscopic features, as described above for the cuprates, occur in the $c$-axis response of these iron arsenides.

To date, only few reports of the $c$-axis response of the iron-arsenides and selenides exist. The available single crystals are typically rather small or cleave very easily making it difficult to obtain ac surfaces of the size and quality required for infrared spectroscopy. To our best knowledge, reports on the $c$-axis conductivity of single crystals exist only for the undoped parent compound Ba-122 and for Ba$_{0.6}$K$_{0.4}$Fe$_2$As$_2$ (BKFA) around optimum doping with $T_c =$ 39~K~\cite{Chen2010PRL,ChengPRB2011}. In both cases a bad-metal-like $c$-axis conductivity with an extremely broad Drude-band was observed in the normal state, with no clear sign of a spin density wave (SDW) or SC gap below $T_N =$ 135~K and $T_c =$ 39~K, respectively. Whereas the absence in the  $c$-axis response of the high energy SDW gap feature occuring in the in-plane response of Ba-122 can be understood in terms of the in-plane oriented wave vector of the SDW order, the lack of a clear SC gap feature in the $c$-axis response of BKFA is rather puzzling since it contrasts with the in-plane response for which an almost complete suppression of the optical conductivity has been observed at $\omega \leq 2\Delta_{SC} \simeq 150\icm$ that is characteristic of a (or several) nearly isotropic gap(s). These data could only be reconciled assuming that the electronic bands and/or Brillouin-zone segments contributing to the in-plane ($c$-axis) transport posses large (small) SC gaps. At present, it is not clear whether calculations of the electronic band-structure support such a scenario. Moreover, to our best knowledge, there exists no information yet on the $c$-axis response of underdoped iron arsenide samples for which evidence of a pseudogap effect in terms of a partial suppression of low-energy states that sets in already well above $T_c$, has been reported from studies of the in-plane response~\cite{Moon2012}. On the other hand, an incoherent $c$-axis response similar to the one of underdoped cuprates has been reported for the iron pnictide compound FeTe$_{0.55}$Se$_{0.45}$ in Ref.~\cite{Moon2011PRL}.

\subsection{$c$-axis response from polycrystalline samples}
Valuable information about the $c$-axis response of layered materials, like the cuprates and iron arsenides, can also be obtained from infrared spectroscopy on polycrystalline samples. This has already been demonstrated for the cuprates for which it was shown that the dominant features in the measured reflectivity spectrum, $R_{poly}$, arise from structures due to the IR-active phonon modes and electronic plasma modes that occur in the $c$-axis reflectivity, $R_c$. Due to the strongly metallic in-plane response, the $ab$-component of the far-infrared reflectivity, $R_{ab}$, is close to unity and exhibits only weak features due to phonons and/or electronic modes and related gap features that hardly show up in $R_{poly}$~\cite{Schlesinger1987}. The close correspondence of the features in $R_{poly}$ with the ones in $R_c$ as measured directly on the $ac$ surface of single crystals has been demonstrated e.g. for LSCO and YBCO~\cite{Bonn1987PRB,Bonn1987PRL,Homes1993PRL,Uchida1996PRB}. Likewise, measurements of polycrystalline Sm,Nd-1111 have already provided information about the SC gap features in the $c$-axis response and have suggested that the optical response of this material is strongly anisotropic~\cite{Dubroka2008PRL}. Notably, this study revealed a clear SC gap feature with an onset of the suppression of the conductivity at $2\Delta_{SC} \simeq$ 300\icm\ in the $c$-axis response. However, the gap-like suppression of $\sigma_{1c}$ is rather gradual and remains incomplete (down to the lowest measured frequency) as compared to the in-plane conductivity for which a much sharper gap feature appears that is characteristic of an isotropic SC gap.

Recently, the new iron arsenide superconductor ACa$_2$Fe$_4$As$_4$F$_2$ (A = Rb, Cs) has become available that is very interesting since it has a bilayer-type structure similar to the one of the cuprates YBCO and Bi-2212 described above~\cite{Wang2016,Wang2017SCM,Wang2018}. Figure~\ref{Fig1}(a) shows that it has pairs of FeAs layers that are separated by Cs or Rb monolayers that are likely more conducting than the insulator-like Ca$_2$F$_2$ double layer that determines the inter-bilayer conductivity. In the following we show that the $c$-axis response of this bilayer-type iron arsenide superconductor, exhibits spectroscopic features that are characteristic of a transverse plasma mode (tPM) and phonon anomalies that can be understood in terms of dynamic charging of the FeAs layers and a corresponding modifications of the local electric fields.

%
%
\section{Experimental method and multilayer model}

\begin{figure}[tb]
\includegraphics[width=\columnwidth]{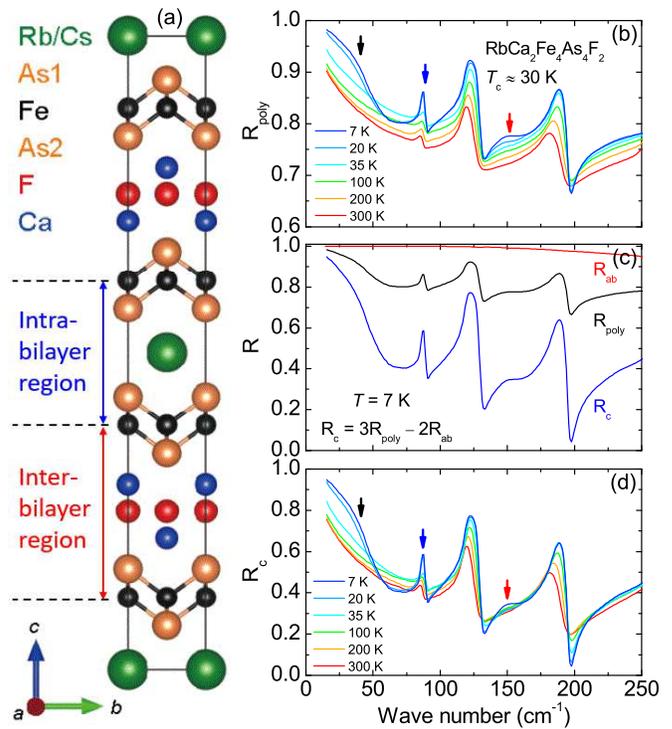}
\caption{ (color online) (a) Sketch of the bilayer-type crystal structure of ACa$_2$Fe$_4$As$_4$F$_2$ (A = Rb, Cs). (b) Far-infrared reflectivity spectra of poly-crystalline RbCa$_2$Fe$_4$As$_4$F$_2$ at several temperatures above and below $T_c \simeq$ 30~K. (c) Comparison of the as measured reflectivity spectra $R_{poly}$ and $R_{ab}$ and the derived $c$-axis spectrum using $R_c = 3R_{poly} - 2R_{ab}$. (d) Temperature dependence of the derived $c$-axis spectra.}
\label{Fig1}
\end{figure}
\begin{figure}[tb]
\includegraphics[width=0.9\columnwidth]{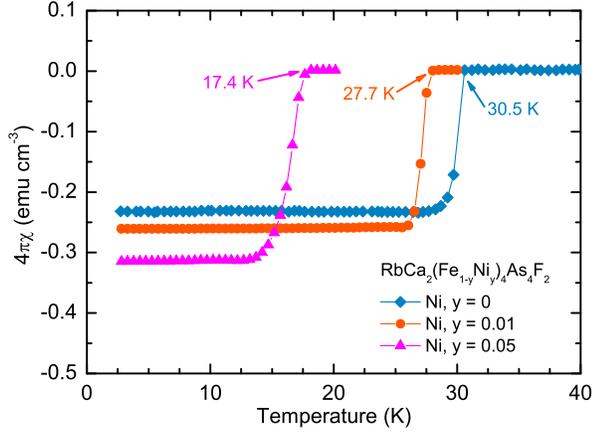}
\caption{ (color online) Temperature dependence of magnetic susceptibility measured at 10~Oe in FC mode for the bilayer iron-based superconductor RbCa$_2$(Fe$_{1-y}$Ni$_{y}$)$_4$As$_4$F$_2$ with $y = $ 0, 0.01 and 0.05.}
\label{Fig2}
\end{figure}

The growth of polycrystalline samples of (Cs,Rb)-12442 and of corresponding single crystals is described in Ref.~\cite{Wang2017SCM}. We followed the same procedure to grow additional Ni substituted polycrystalline samples with 1\% and 5\% of Ni for Fe. The superconducting transition temperatures of these polycrystalline samples of $T_c \simeq$ 30~K, 27~K and 17~K for 0\%, 1\% and 5\% of Ni, respectively, as obtained from the dc magnetisation data are shown in Fig.~\ref{Fig2}.

The infrared reflectivity study of the in-plane response of a Cs-12442 single crystal is presented in Ref.~\cite{Xu2019PRB}. We used the same optical setup to obtain the far-infrared reflectivity spectra of the corresponding polycrystalline samples of Rb-12442 with 0\%, 1\% and 5\% of Ni. The data were obtained on surfaces of sintered pellets that were polished to optical grade using oil-based diamond paste.

Figure~\ref{Fig1}(a) shows a sketch of the bilayer-type crystal structure of the 12442 compound with the intra-bilayer and the inter-bilayer regions indicated by blue and red arrows. Figs.~\ref{Fig1}(b) to \ref{Fig1}(d) illustrate the geometrical averaging approach that was used to deduce from the measured reflectivity spectra of the polycrystalline sample, $R_{poly}$, and the in-plane reflectivity spectra of the single crystal, $R_{ab}$, the $c$-axis component of the reflectivity, according to: $R_{poly} = (1/3)R_c + (2/3)R_{ab}$. Note that this approach is only valid if the size of the crystallites exceeds the effective wave length of the light, $\lambda/n$, where $\lambda$ is the wavelength in vacuum and $n$ the refractive index of the material~\cite{Sihvola1999,Mayerhofer2005}. Thanks to the very large values of $n$ (especially for the $ab$-component), this condition can be fulfilled for micrometer-sized crystallites, even in the FIR region with $\lambda \sim$ 100~$\mu$m. Alternatively, we used an effective medium approximation~\cite{Sihvola1999} to analyse the data as shown in the Appendix~\ref{apppendixA}. While both approaches yielded qualitatively similar results, we found that the $R_c$ spectra obtained from the geometrical averaging were generally in better agreement with the multilayer model. Accordingly, we focus the following discussion on the data obtained with the geometrical averaging approach.

The further analysis of the obtained $R_c$ spectra in terms of the dielectric function $\varepsilon_c$ with $R_c = |\frac{1- \sqrt{\varepsilon_c}}{1 + \sqrt{\varepsilon_c}}|^2$ and $\varepsilon_c = \varepsilon + \varepsilon_{ph}$, where $\varepsilon$ and $\varepsilon_{ph}$ account for the electronic and phonon contributions, respectively, has been performed with the same multilayer model that was  previously used for the $c$-axis spectra of single crystals of the bilayer cuprates YBCO and Bi-2212~\cite{Dubroka2010,Dubroka2011PRL}. The formula for the electronic dielectric function, $\varepsilon$, reads,
\begin{equation}
\label{eq_c_axis}
\varepsilon(\omega) = \frac{\varepsilon_{\mathrm{bl}}(\omega)\varepsilon_{\mathrm{int}}(\omega)}{z_{\mathrm{bl}}\varepsilon_{\mathrm{int}}(\omega)+z_{\mathrm{int}}\varepsilon_{\mathrm{bl}}(\omega)}, \end{equation}
where $\varepsilon_{\mathrm{bl}}(\omega)$ and $\varepsilon_{\mathrm{int}}(\omega)$ are the local dielectric functions connected to the two different local conductivities as $\varepsilon_{\mathrm{bl}}(\omega) = \varepsilon_\infty + \frac{i\sigma^{\mathrm{bl}}(\omega)}{\omega\epsilon_0}$ and $\varepsilon_{\mathrm{int}}(\omega) = \varepsilon_\infty + \frac{i\sigma^{\mathrm{int}}(\omega)}{\omega\epsilon_0}$, and the factors $z_{\mathrm{bl}} = d_{\mathrm{bl}}/(d_{\mathrm{bl}} + d_{\mathrm{int}})$ and $z_{\mathrm{int}} = d_{\mathrm{int}}/(d_{\mathrm{bl}} + d_{\mathrm{int}})$ are the volume fractions of the intra- and inter-bilayer regions, respectively. The local dielectric functions $\varepsilon_{\mathrm{bl}}(\omega)$ and $\varepsilon_{\mathrm{int}}(\omega)$ are modeled by the following set of Drude-Lorentz oscillators,
\begin{eqnarray}
\label{eq_bl_int}
\varepsilon_{a}(\omega) &=& \varepsilon_{\infty} - \frac{\Omega^2_{pS,a}}{\omega(\omega+i\delta)} - \frac{\Omega^2_{pD,a}}{\omega^2 + i\omega\gamma_{D,a}} \nonumber \\
& &+ \frac{S^2_{a}}{\omega^2_{0,a} - \omega^2 - i\omega\Gamma_{a}}, a \in \{\mathrm{bl},\mathrm{int}\}.
\end{eqnarray}
The second and third terms account for the response of the superconducting condensate and the quasiparticles with finite scattering rate (Drude term), respectively. The fourth term represents an additional Lorentzian oscillator that describes the higher energy part of the spectra.

%
%
\section{Results and Discussion}
\begin{figure}[tb]
\includegraphics[width=\columnwidth]{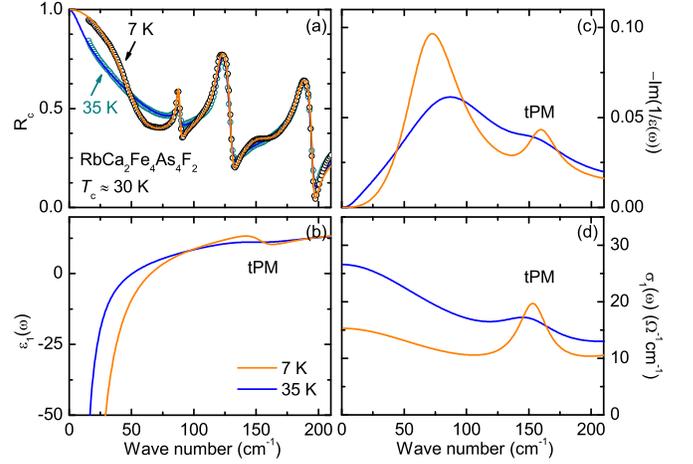}
\caption{ (color online) (a) Fits of the $c$-axis reflectivity spectra with the multilayer model for RbCa$_2$Fe$_4$As$_4$F$_2$ at 35~K just above $T_c$ and at 7~K well below $T_c$. (b) Real part of the dielectric function, (c) loss function, and (d) optical conductivity at 35 and 7~K after, subtraction of the sharp phonon modes.
}
\label{Fig3}
\end{figure}
The most important features in the $c$-axis response of Rb-12442 are indicated with coloured arrows in Figs.~\ref{Fig1}(b) and ~\ref{Fig1}(d). The red arrow marks the transverse plasma mode (tPM) around 150\icm\ that develops below $T_c \simeq$ 30~K and arises from the out-of-phase electronic response of the intra- and inter-bilayer regions (as described in the introduction). The blue arrow highlights an IR-active phonon mode that exhibits a pronounced SC-induced anomaly in terms of a sudden increase of its spectral weight below $T_c$. The low frequency of this phonon mode of about 90\icm\ suggests that it involves the heavy Rb atoms in the centre of the intra-bilayer unit (see Fig.~\ref{Fig1}(a)). Its displacement pattern (eigenvector) is thus likely similar to that of the Y-mode at 190\icm\ in YBCO. Notably, this latter Y-mode exhibits a similar SC-induced anomaly with a sudden increase of its oscillator strength below $T_c$ as was shown in Refs.~\cite{Schutzmann1995PRB,Homes1995CJP,Bernhard2000PRB}. Finally, the black arrow marks the low-frequency plasma edge due to the lowest longitudinal plasma mode, that arises from a weak, Josephson-type coupling of the FeAs layers through the inter-bilayer regions, i.e. the insulating Ca$_2$F$_2$ layers. Notably, the counterparts of all three anomalous low temperature features were previously observed in the $c$-axis response of underdoped YBCO.

Figure~\ref{Fig3}(a) shows the $R_c$ spectra (open symbols) at $T =$ 35~K in the normal state just above $T_c$ and at 7~K $\ll T_c$, together with the best fits obtained with the multilayer model (solid lines). It demonstrates that the multilayer model provides a good description of the $c$-axis response obtained from a polycrystalline sample. Note that the contribution of the infrared-active phonon modes to the dielectric function has been fitted using the following function,
\begin{equation}
\label{eq_ph}
\varepsilon_{ph}(\omega) = \sum_{ph} \frac{\Omega^2_{ph} - i\omega\beta_{ph}}{\omega^2_{0,ph} - \omega^2 - i\omega\gamma_{ph}},
\end{equation}
that has been added to the expression on the right hand side of Eq.~\ref{eq_c_axis}. The sum in Eq.~\ref{eq_ph} consists of Lorentzian functions, each with a resonance frequency $\omega_{0,ph}$, a line width $\gamma_{ph}$ and a complex oscillator strength $\Omega^2_{ph} - i\omega\beta_{ph}$. The imaginary part of the oscillator strength $i\omega\beta_{ph}$ enables us to describe the lineshape of the phonons even if they are asymmetric because they are renormalized by an interaction, for example, with other phonons or the electronic background.

The electronic $c$-axis response functions obtained from the multilayer modelling, after the subtraction of the IR-active phonon modes, are displayed in Figs.~\ref{Fig3}(b) to \ref{Fig3}(d),  in terms of the real part of the dielectric function, $\varepsilon_1(\omega)$, the loss-function, $-\mathrm{Im}(1/\varepsilon(\omega))$, and the real part of the conductivity, $\sigma_1(\omega)$, respectively. The normal state spectra at 35~K reveal a rather weak Drude-response with the screened plasma frequency of $\omega_{p,n}/\sqrt{\varepsilon_\infty} \approx 70\icm$ (with $\varepsilon_\infty = 4$ fixed in the multilayer modelling) and only a very faint and broad peak around 150\icm\ due to a strongly overdamped tPM. In the SC state at 7~K, the spectral weight of the Drude peak is strongly reduced and the missing weight is redistributed to a $\delta$-function at zero frequency and to the peak around 150\icm\ which arises from a strongly enhanced and sharpened tPM.

\begin{figure}[tb]
\includegraphics[width=\columnwidth]{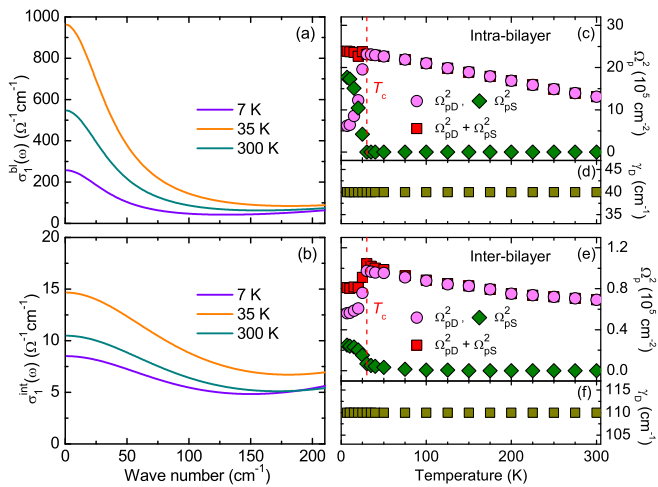}
\caption{ (color online) The local conductivities of (a) the intra-bilayer region $\sigma_1^{\mathrm{bl}}(\omega)$ and (b) the inter-bilayer region $\sigma_1^{\mathrm{int}}(\omega)$ at 300, 35 and 7~K. Temperature dependences of the obtained values of the fitting parameters, the squared plasma frequency and broadening of the Drude term, $\Omega^2_{pD}$ and $\gamma_D$, and the squared plasma frequency of the SC condensate, $\Omega^2_{pS}$, that is proportional to the spectral weight of the SC $\delta$-function for (c)--(d) the intra-bilayer and (e)--(f) the inter-bilayer response, respectively. The $\gamma_D$ values of 40\icm\ and 110\icm, respectively, have been obtained by fitting the 7~K data and kept fixed for higher temperatures.
}
\label{Fig4}
\end{figure}
%
\begin{table}[tb]
\caption{\label{Table1}%
Values of the parameters $\Omega_{pD}$, $\gamma_D$ and $\Omega_{pS}$ entering Eq.~\ref{eq_bl_int} obtained by fitting the $c$-axis reflectivity spectra of of RbCa$_2$Fe$_4$As$_4$F$_2$ for three representative temperatures. All quantities are in units of \icm.}
\begin{ruledtabular}
\begin{tabular}{ccccccc}
& \multicolumn{3}{c}{intra-bilayer} & \multicolumn{3}{c}{inter-bilayer} \\
\cline{2-4}\cline{5-7}
$T$ (K) & $\Omega_{pD}$  & $\gamma_D$ & $\Omega_{pS}$  &$\Omega_{pD}$  & $\gamma_D$ & $\Omega_{pS}$ \\
\hline
300  & 786 & 40 & 1330 & 237 & 110 & 158\\
35	 & 1520	& 40  & 0 & 311 & 110 & 72 \\
7    & 1146 & 40 & 0 & 263 & 110 & 0\\
\end{tabular}
\end{ruledtabular}
\end{table}
Figures~\ref{Fig4}(a) and \ref{Fig4}(b) show the spectra of the local conductivities of the intra-bilayer and the inter-bilayer regions, $\sigma_1^{\mathrm{bl}}(\omega)$ and $\sigma_1^{\mathrm{int}}(\omega)$, respectively, as obtained from the multilayer fitting using Eq.~\ref{eq_bl_int}. The frequencies of the Lorentzian oscillators were fixed at about 1000\icm\ and the oscillators are almost temperature independent. The temperature dependencies of the obtained fitting parameters, the plasma frequency and the width of the Drude-term, $\Omega_{pD}^2$ and $\gamma_D$, as well as the weight of the SC $\delta$-function, $\Omega_{pS}^2$ are displayed in Figs.~\ref{Fig4}(c--d) and~\ref{Fig4}(e-f) for the intra-bilayer and the inter-bilayer response, respectively. The values of $\gamma_D$ of 40\icm\ and 110\icm\ have been obtained by fitting the 7~K data and kept fixed for higher
temperatures. The values of the parameters at several representative temperatures are summarized in Table~\ref{Table1}.

At 35~K in the normal state just above $T_c$, this multilayer analysis reveals a rather large, more than 20-fold difference between the squared plasma frequencies of the intra- and the inter-bilayer regions (recall that $\Omega_{pD}^2 \sim n/m^\ast$, where $n$ and $m^\ast$ are the concentration and effective mass of the carriers). Interestingly, the absolute values of $\sigma_1^{\mathrm{bl}}(\omega)$ and $\sigma_1^{\mathrm{int}}(\omega)$ are similar to the ones reported for underdoped YBCO with $T_c \sim$ 60~K~\cite{Dubroka2011PRL,Dubroka2010}.

An important difference concerns the spectral shape of the inter-bilayer conductivity, $\sigma_1^{\mathrm{int}}(\omega)$, which in underdoped YBCO was found to be insulator-like with a steady decrease of the absolute values as temperature is reduced (see Fig.~1(d) of Ref.~\cite{Dubroka2011PRL}). The insulator-like behaviour of $\sigma_1^{\mathrm{int}}(\omega)$ was interpreted in terms of the aforementioned pseudogap phenomenon which depletes the low-energy electronic states in the antinodal parts of the Fermi-surface (near to (0,$\pi$) or ($\pi$,0)) at which the $c$-axis hopping parameter across the inter-bilayer region, $t_\perp^{\mathrm{int}}$, is maximal~\cite{Dubroka2011PRL}.

In contrast, for Rb-12442 the inter-bilayer conductivity $\sigma_1^{\mathrm{int}}(\omega)$ is well described in terms of a Drude-type response of itinerant carriers. Moreover, $\sigma_1^{\mathrm{int}}(\omega)$ has a metallic temperature dependence and exhibits a nearly two-fold increase between 300 and 35~K. It is quite remarkable that the inter-bilayer response remains coherent, despite of the very small absolute value of $\sigma_1^{\mathrm{int}}(\omega \rightarrow 0, \mathrm{35~K}) \simeq$ 15~$\Omega^{-1}$\icm. Certainly, the inter-bilayer response of the Rb-12442 sample shows no sign of a pseudogap-effect, similar to the one in the $c$-axis response of the underdoped cuprates.

\begin{figure}[tb]
\includegraphics[width=\columnwidth]{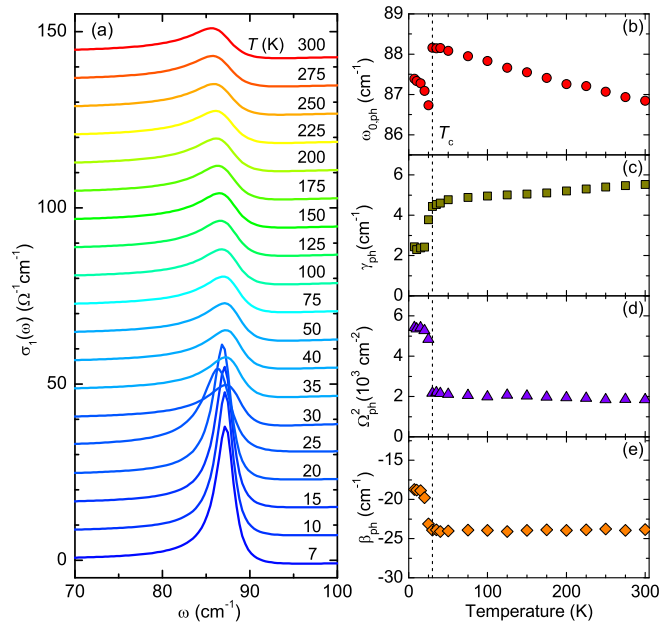}
\caption{ (color online) (a) Temperature dependence of the lineshape of the Rb phonon mode. (b)--(e) Temperature dependences of the resonance frequency $\omega_{0,ph}$, the line width $\gamma_{ph}$, the oscillator strength $\Omega_{ph}^2$, and the parameter $\beta_{ph}$ of the phonon as obtained from a fit with an asymmetric Lorentz-function. The vertical dashed line denotes the superconducting transition temperature $T_c$.}
\label{Fig5}
\end{figure}

Concerning the SC-induced changes at $T \ll T_c \simeq$ 30~K, it is evident from Figs.~\ref{Fig4}(a) and \ref{Fig4}(b) that both $\sigma_1^{\mathrm{bl}}(\omega)$ and $\sigma_1^{\mathrm{int}}(\omega)$ exhibit a clear gap-like suppression of the regular part of the low energy response, i.e. the Drude-peak is strongly reduced and the missing spectral weight is mostly transferred to the $\delta$-function at zero frequency that represents the loss-free response of the SC condensate. For the intra-bilayer response, $\sigma_1^{\mathrm{bl}}(\omega)$, the spectral weight loss of the Drude-peak amounts to about 70\% and is fully balanced by the corresponding spectral weight gain of the SC condensate, as shown in Fig.~\ref{Fig4}(c).  For the inter-bilayer conductivity, $\sigma_1^{\mathrm{int}}(\omega)$, the spectral weight loss of the Drude-peak is only about 40\% and, as shown in Fig.~\ref{Fig4}(e), it is not fully compensated by the spectral weight of the $\delta$-function at zero frequency. Some of the missing spectral weight could be transferred to the infrared-active phonon modes, which are not explicitly included in the multilayer modelling. Figure~\ref{Fig5} shows that the low-frequency mode at 90\icm\ exhibits indeed a significant spectral weight increase below $T_c$. Taking into account also a similar spectral weight gain below $T_c$ of the mode at 120\icm\ (data are not shown), we estimate that this can account for about half of the missing spectral weight below  $T_c$ of $\sigma_1^{\mathrm{int}}(\omega)$. The remaining inconsistency is likely due to the uncertainty of fitting the contribution of the SC condensate to  $\sigma_1^{\mathrm{int}}(\omega)$ which strongly depends on the low frequency part of the experimental data ($\omega <$ 50\icm) for which the error bars are the largest.

Overall, the $c$-axis response of this bilayer compound exhibits quite a sizeable suppression of the Drude-response due to the SC gap formation. The suppression of the low-frequency conductivity is not quite as strong as for the in-plane response of a single crystal, for which the gap(s) appear(s) to be almost isotropic and complete with almost zero conductivity at $\omega < 2\Delta \approx 110\icm$~\cite{Xu2019PRB}. Nevertheless, it is much stronger than the one reported for the $c$-axis response of an optimally doped BKFA single crystal~\cite{ChengPRB2011} where no clear sign of a SC gap was observed. Note that the weaker signature of the SC gap in the inter-bilayer response can be explained provided that the band that is most dispersive along the $c$-axis direction has a smaller SC gap than the bands that are more dispersive along the in-plane direction. This conjecture agrees with the ARPES data in Fig.~2 of Ref.~\cite{Wu2020} which reveal that the $\alpha$ and $\beta$ bands with larger gaps of 2$\Delta \approx$ 8--16~meV have a more pronounced in-plane dispersion than the $\gamma$ band with a smaller gap 2$\Delta \approx$ 4~meV that has the strongest $c$-axis dispersion.

Next, we discuss the SC-induced anomaly of the Rb-phonon mode around 90\icm\ that is shown in Fig.~\ref{Fig5}(a). The corresponding temperature dependences of the parameters obtained from fitting the $R_c$ spectra with Eq.~\ref{eq_ph} are displayed in Figs.~\ref{Fig5}(b) to \ref{Fig5}(e). All parameters exhibit clear anomalies that start right below $T_c \simeq$ 30~K. It was previously shown for the underdoped cuprates, that the phonon anomalies are particularly sensitive to the onset of local SC pairing interactions at $T > T_c$~\cite{Dubroka2011PRL}. The sharp onset of the anomaly of the Rb-mode right below $T_c \simeq$ 30~K, is therefore a clear indication that such a precursor SC pairing does not occur in Rb-12442. This conclusion is supported by the equally sharp anomaly right below $T_c$ of the tPM and the related $\delta$-function at zero frequency in $\sigma_1^{\mathrm{bl}}$ [see Figs.~\ref{Fig4}(a) and \ref{Fig4}(c)]. As already discussed in the previous paragraph, the multilayer-fitting of the $\delta$-function in $\sigma_1^{\mathrm{int}}$ has a large uncertainty, its very weak onset slightly above $T_c$ is therefore likely an artefact.

In the following, we address possible connections between the tPM feature and the bilayer splitting, i.e. the splitting between the bonding and antibonding bands of the bilayer unit. For the cuprates it was recently shown, based on the multilayer modelling of the $c$-axis response, that such a bilayer splitting is detectable above a hole doping of $p \approx 0.10$, where it increases in magnitude towards optimum doping up to 30--60~meV~\cite{Mallett2019PRB,Dordevic2004}. Specifically, it was shown that significantly better fits are obtained if $\sigma_1^{\mathrm{bl}}(\omega)$ is described by a Lorentzian-function with a peak at an energy comparable to the magnitude of the bilayer splitting, rather than by a Drude-response with a peak at zero energy. In the normal state, this Lorentzian peak accounts for the interband transition between the bonding and anti-bonding bands of the CuO$_2$ bilayer unit that is strongly damped. In the superconducting state, it corresponds to a collective mode that is only weakly damped if the gap 2$\Delta$ is larger than the bilayer-splitting~\cite{Chaloupka2009}. Otherwise, if 2$\Delta$ is smaller than the bilayer-splitting, this mode is also strongly damped and the tPM feature and the phonon anomalies due to the related local electric field effects are expected to be weak. For the present Rb-12442 sample, we find no evidence of such a bilayer splitting of an energy scale larger than about 5~meV, i.e. we verified that describing $\sigma_1^{\mathrm{bl}}(\omega)$  by a Lorentzian-function with a peak energy above 5~meV (instead of the Drude-function shown in Fig.~\ref{Fig4}) does not improve the quality of the multilayer fitting. Note that these multilayer fits are not expected to be sensitive to a bilayer splitting smaller than the width of the Drude-function of $\sim$ 5~meV. A similarly small bilayer splitting has indeed been observed in a very recent ARPES study of a K-12442 crystal which has the same type of bilayer structure~\cite{Wu2020}. As shown in Fig. 2 of \cite{Wu2020}, a clear bilayer splitting is observed for the hole-like $\gamma$ and $\beta$ bands near the Brillouin-zone center. The largest splitting near the Fermi-surface occurs for the $\gamma$ band for which it is in the range from 5 to 10~meV, slightly larger than the reported value of the SC gap of the $\gamma$ band of 2$\Delta \approx$ 4~meV~\cite{Wu2020}. This raises the question of how it is possible that a reasonably well defined tPM appears in the spectra. A likely scenario is the following: The formation of relatively large gaps on the $\alpha$ and $\beta$ bands (2$\Delta \approx$ 8--16~meV) leads to a dramatic sharpening of the bilayer splitted $\gamma$ bands below $T_c$. This effect can indeed be seen in Fig.~3 of Ref. ~\cite{Wu2020}. As a consequence, the corresponding low-energy optical transition also sharpens considerably giving rise to a clear tPM feature. The very low energy of the transition allows one to use the phenomenological multilayer model in its simplest form, when analyzing the data, as we did in the previous paragraphs

\begin{figure}[tb]
\includegraphics[width=0.9\columnwidth]{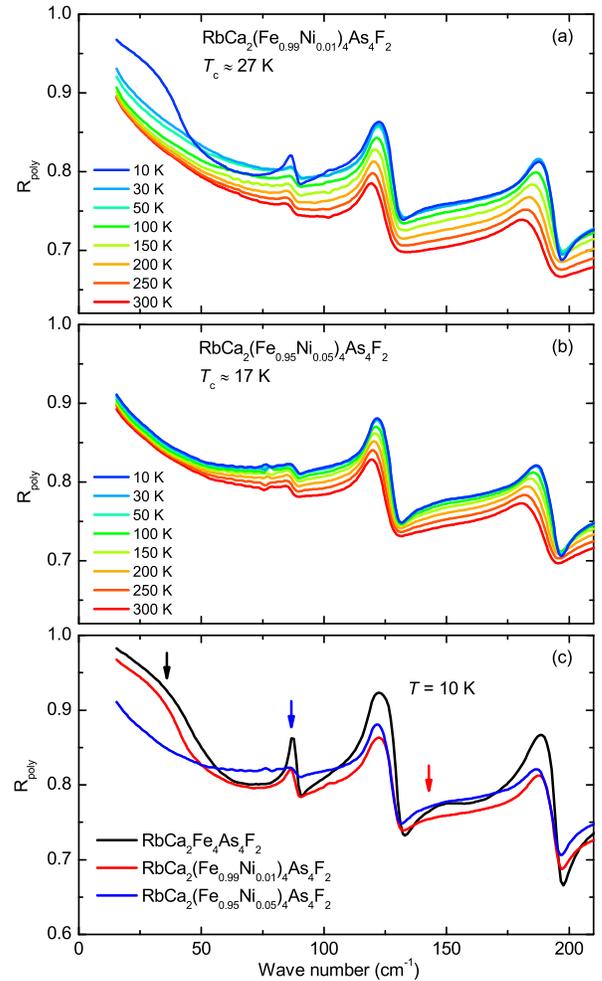}
\caption{ (color online) (a)--(b) Temperature dependent far-infrared reflectivity spectra of poly-crystalline RbCa$_2$(Fe$_{(1-y)}$Ni$_{y}$)$_4$As$_4$F$_2$ with $y =$ 0.01 and 0.05. (c) Comparison of the reflectivity spectra at 10~K of RbCa$_2$(Fe$_{(1-y)}$Ni$_{y}$)$_4$As$_4$F$_2$ with $y =$ 0, 0.01 and 0.05.}
\label{Fig6}
\end{figure}
Another important issues is that Rb-12442 is expected to be strongly hole doped and located on the overdoped rather than on the underdoped side of the phase diagram. The hole doping of the FeAs layers of Rb-12442 is nominally the same as in Ba$_{0.5}$K$_{0.5}$Fe$_2$As$_2$~\cite{Rotter2008,Avci2012,Boehmer2015}. Accordingly, it is not too surprising that there is no pseudogap effect above $T_c$ due to  competing electronic and/or magnetic orders and no onset of local SC pairing at a temperature significantly higher than $T_c$.

\begin{figure}[tb]
\includegraphics[width=\columnwidth]{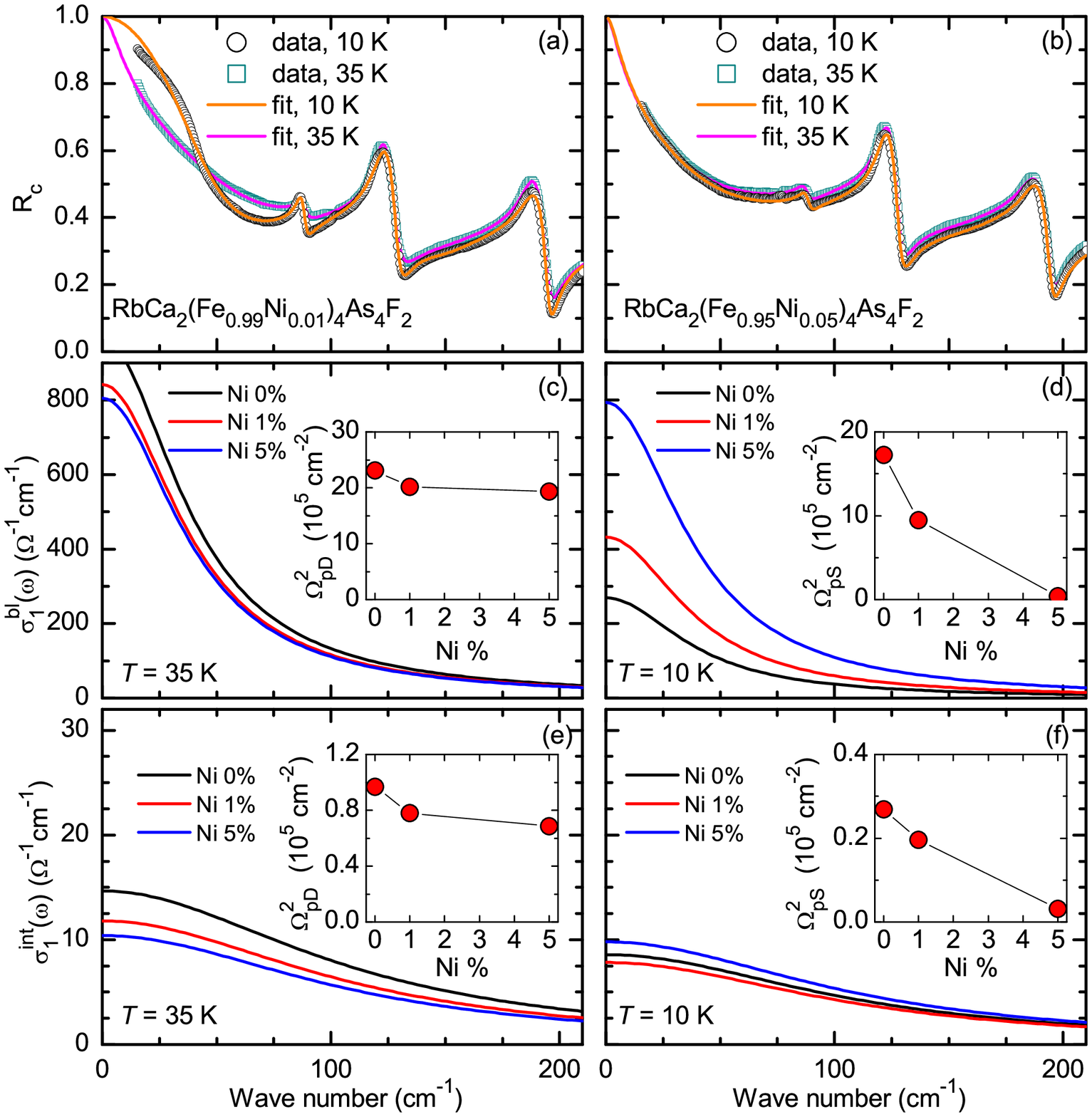}
\caption{ (color online) (a--b) Fits of the $c$-axis reflectivity spectra with the multilayer model for RbCa$_2$(Fe$_{(1-y)}$Ni$_{y}$)$_4$As$_4$F$_2$ with $y =$ 0.01 and 0.05 at 35~K just above $T_c$ and 7~K well below $T_c$. (c)--(f) Comparison of the local conductivities of the intra-bilayer region $\sigma_1^{\mathrm{bl}}(\omega)$ and the inter-bilayer region $\sigma_1^{\mathrm{int}}(\omega)$ for RbCa$_2$(Fe$_{(1-y)}$Ni$_{y}$)$_4$As$_4$F$_2$ with $y =$ 0, 0.01 and 0.05 at representative temperatures of 35~K and 7~K. The insets show the evolution of the corresponding normal state and SC plasma frequencies as a function of the Ni-concentration.
}
\label{Fig7}
\end{figure}
Accordingly, we have also studied a series of polycrystalline, Ni-substituted samples  RbCa$_2$(Fe$_{(1-y)}$Ni$_{y}$)$_4$As$_4$F$_2$ with $y =$ 0.01 and 0.05 for which the hole doping is progressively reduced toward the underdoped regime. It was previously shown in BaFe$_{2-y}$Ni$_{2y}$As$_2$ that Ni substitution gives rise to electron doping at a rate that is about twice faster than for Co substitution ~\cite{Ni2008PRB,Canfield2009PRB,Ni2010PRB}. It was also shown for Ba$_{1-x}$K$_x$Fe$_{2-y}$(Co,Ni)$_y$As$_2$ that co-doping with Co or Ni counteracts the hole doping due to the K-substitution~\cite{Goltz2014PRB}. Moreover, for CaKFe$_{4-4y}$Ni$_{4y}$As$_4$, which at $y = 0$ has the same nominal hole doping as Rb-12442, it has already been demonstrated that Ni substitution allows one to reach the underdoped side of the phase diagram and induce a magnetic order (and possibly also nematic and orbital orders) that coexists and competes with SC~\cite{Meier2018,Ding2018}.

The magnetisation data of our polycrystalline samples in Fig.~\ref{Fig2} confirm that the Ni-substitution gives rise to a strong suppression of the SC transition temperature with $T_c$ values of 30, 27 and 17~K for $y =$ 0, 0.01 and 0.05, respectively. Figures~\ref{Fig6}(a) and \ref{Fig6}(b) show the temperature dependence of the far-infrared spectra of the as measured reflectivity, $R_{poly}$, for the polycrystalline samples with $y =$ 0.01 and 0.05, respectively. Figure ~\ref{Fig6}(c) compares the spectra at $T = \mathrm{10 K} < T_c$ for the series with $y =$ 0, 0.01 and 0.05. The characteristic superconductivity-induced features due to the tPM around 150\icm, the anomaly of the Rb-phonon mode around 90\icm, and the low-frequency plasma edge around 40\icm\ (that is determined by $\Omega_{pS}^{\mathrm{int}}$ ) are marked by the same kind of arrows as in Fig.~\ref{Fig1}(b). Figure~\ref{Fig6} reveals that these SC-induced anomalies are significantly weakened for the sample with $y =$ 0.01 and almost absent for $y =$ 0.05. Moreover, Figs.~\ref{Fig6}(a) and \ref{Fig6}(b) establish that the Ni-substitution does not give rise to any anomalous changes of the spectra in the normal state above $T_c$, that could be associated either with a pseudogap effect due to a competing order, or with precursor pairing correlations above $T_c$.

Figures~\ref{Fig7}(a) and \ref{Fig7}(b) show the FIR spectra of $R_c$ at 35 and 10~K (open symbols) at $y =$ 0.01 and 0.05, respectively, that have been derived using the same geometrical averaging as discussed above and shown in Fig.~\ref{Fig1} for $y = 0$. For the in-plane spectra we used the data of the undoped single crystal from Ref.~\cite{Xu2019PRB}. The solid lines represent the best fits obtained with the multilayer model. Finally, Figs.~\ref{Fig7}(c)--\ref{Fig7}(f) show a comparison of the local conductivities $\sigma_1^{\mathrm{bl}}$ and $\sigma_1^{\mathrm{int}}$ at 35 and 10~K. The insets show the evolution of the corresponding normal state and SC plasma frequencies as a function of the Ni-content. They reveal that the Ni substitution hardly affects the normal state conductivity and the related plasma frequency $\Omega_{pD}$, whereas it gives rise to a strong suppression of the SC $\delta$-function and $\Omega_{pS}$.

%
%
\section{Summary and Conclusion}
With infrared spectroscopy we have studied RbCa$_2$(Fe$_{1-y}$Ni$_y$)$_4$As$_4$F$_2$ (Rb-12442) with $y =$ 0, 0.01, and 0.05 which has a bilayer-type structure similar to the cuprates, like YBa$_2$Cu$_3$O$_7$ (YBCO) or Bi2212. From combined measurements of polycrystalline samples and the in-plane response of corresponding single crystals, and using a geometrical averaging approach, we have derived the $c$-axis conductivity of these samples. In analogy to the cuprates, the $c$-axis response exhibits several spectroscopic features that are characteristic of a layered structure with very different strengths of the electronic response of the intra- and inter-bilayer regions. In particular, we find that a transverse plasma mode (tPM) develops around 150\icm\ that is strongly overdamped in the normal state but sharpens and is strongly enhanced below $T_c \simeq$ 30~K. Along with the tPM comes a pronounced SC-induced anomaly of the lowest IR-active phonon mode around 90\icm\ that involves the displacement of the heavy Rb ions and arises from the local-electric-field effects caused by the dynamical charging of the FeAs layers connected with the formation of the tPM. Moreover, we observe a pronounced low-frequency plasma edge (around 50\icm) that is characteristic of a very weak electronic inter-bilayer response. Using a multilayer model, similar to the one previously used for the $c$-axis response of underdoped YBCO, we achieved a good description of the experimental data. The obtained values of the inter- and intra-bilayer conductivities are comparable to the ones of underdoped YBCO with $T_c =$ 60~K. A marked difference concerns the absence of a pseudogap effect in the inter-bilayer conductivity of Rb-12442, which, despite of a very small absolute value of $\sigma_1^{\mathrm{int}}(\omega \rightarrow 0) \simeq$ 15 $\Omega^{-1}\mathrm{cm}^{-1}$, remains metal-like. Moreover, from the temperature dependence of the tPM and the anomaly of the Rb-phonon mode, we obtain no evidence for the onset of precursor SC pairing correlations well above the bulk SC transition temperature, $T_c$, at which the phase coherence becomes macroscopic. Such an onset is also not observed for the Ni-substituted sample for which the hole doping is reduced toward the underdoped regime. This is despite of a strong Ni-induced suppression of the plasma frequency of the SC $\delta$-function.

Our observations suggest that the pseudogap and the onset of local superconducting pairing correlations well above $T_c$ are unique features of the cuprates, i.e. they are not just the result of a strong electronic anisotropy due to a very weak interlayer conductivity. The multilayer analysis also reveals clear signatures of the SC gap formation in the inter- and intra-bilayer conductivities that are much more pronounced that in the previous report on a Ba$_{0.6}$K$_{0.4}$Fe$_2$As$_2$ (BKFA) crystal without such a bilayer structure~\cite{ChengPRB2011}. A sizeable, ungapped part of the low-frequency conductivity, that is more pronounced (related to the overall strength of the response) for the inter-bilayer than for the intra-bilayer conductivity, can be understood in terms of the multi-band structure with a $\gamma$ band that has a rather small gap and is more dispersive along the $c$-axis than the other bands with larger gaps that disperse more strongly in the $k_x - k_y$ plane. Such different dispersion behaviours and gap magnitudes have indeed been recently reported in an ARPES study of an isotructural K-12442 single crystal~\cite{Wu2020}.

%
%
\section{Acknowledgments}
Work at the University of Fribourg was supported by the Schweizer Nationalfonds (SNF) by Grant No. 200020-172611. D.M. and A.D were supported by the MEYS of the Czech Republic under the project CEITEC 2020 (LQ1601), and by the Czech Science Foundation (GA\v{C}R) under Project No. GA20-10377S.

%
\appendix

\section{Effective medium approximation}
\label{apppendixA}

According to the Maxwell-Garnett theory, the effective dielectric properties ($\varepsilon_{\mathrm{eff}}$) of a poly-crystalline material is determined by the following equation:
\begin{equation}
\varepsilon_{\mathrm{eff}} = \varepsilon_{ab} \frac{(\varepsilon_{c}+2\varepsilon_{ab}) + 2p(\varepsilon_{c}-\varepsilon_{ab})}{(\varepsilon_{c}+2\varepsilon_{ab}) - p(\varepsilon_{c}-\varepsilon_{ab})},
\end{equation}
where the effective medium consists of a matrix with $\varepsilon_{ab}$ ($ab$-plane response) and inclusions with $\varepsilon_{c}$ ($c$-axis response), and $p$ corresponds to the volume fraction of the inclusions in the effective medium.

\begin{figure}[tbh]
\includegraphics[width=\columnwidth]{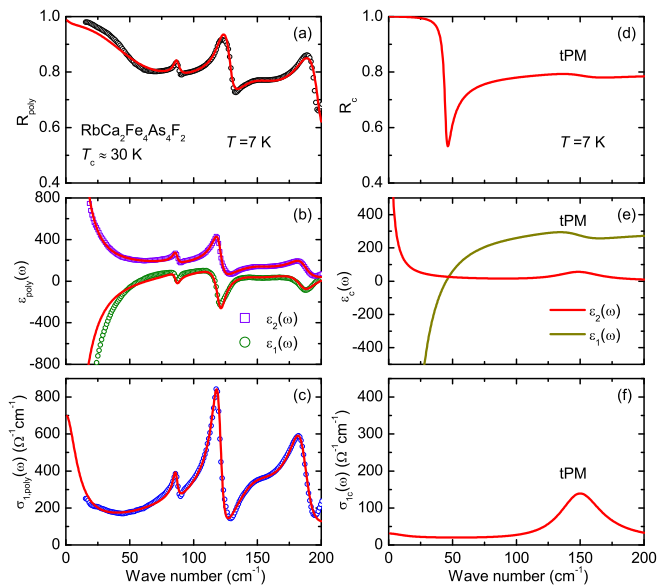}
\caption{(Color online) Fits of (a) the reflectivity, (b) the dielectric function, and (c) the real part of optical conductivity spectra with effective medium approximation for poly-crystalline RbCa$_2$Fe$_4$As$_4$F$_2$ at 7~K. The $c$-axis response of (d) the reflectivity, (e) the dielectric function, and (f) the real part of optical conductivity at 7~K after subtraction of the sharp phonon modes.}
\label{Fig8}
\end{figure}
The effective medium approximation dielectric function is fit to the data of poly-crystalline RbCa$_2$Fe$_4$As$_4$F$_2$, where $\varepsilon_{ab}$ is described by a simple Drude model, and $\varepsilon_{c}$ by the multilayer model mentioned in the main text, and the volume fraction $p = 1/3$. Figs.~\ref{Fig8}(a)--\ref{Fig8}(c) show the fitting results to the reflectivity, the dielectric function, and the real part of optical conductivity of poly-crystalline RbCa$_2$Fe$_4$As$_4$F$_2$ at 7~K, respectively. The corresponding electronic $c$-axis response functions obtained from the effective medium approximation, after the subtraction of the IR-active phonon modes, are displayed in Figs.~\ref{Fig8}(d) to \ref{Fig8}(f), in terms of the reflectivity $R_c$, the dielectric function $\varepsilon_c(\omega)$, and the real part of the conductivity $\sigma_{1c}(\omega)$, respectively. Qualitatively, the effective medium approximation approach yielded similar results to the geometrical averaging approach we discussed in the main text.

%
%

\end{document}